\documentclass[12pt]{article}
\usepackage{amsmath,amssymb,amsthm,amssymb,graphicx}
\title{Complete account of randomness in the EPR-Bohm-Bell experiment}
\author{David Avis\\
Computer Science, McGill University\\
 3480 University Montreal, Quebec, Canada H3A 2A7\\
Paul Fischer\\
Informatics and Mathematical Modelling\\
Technical University of Denmark\\
DK-2800  Kgs. Lyngby, Denmark\\[2mm]
Astrid Hilbert and Andrei Khrennikov\\
International Center for Mathematical Modeling
\\in Physics and Cognitive Sciences,\\
University of V\"axj\"o, S-35195, Sweden}

\begin{document}

\maketitle

\begin{abstract}We show that paradoxical consequences of violations
of Bell's inequality are induced by the use of an unsuitable
probabilistic description for the EPR-Bohm-Bell experiment. The
conventional description (due to Bell) is based on a 
combination of statistical data collected for different settings
of polarization beam splitters (PBSs). 
In fact, such data consists of some conditional probabilities
which only partially define a probability space.
Ignoring this conditioning leads to
apparent contradictions in the classical
probabilistic model (due to Kolmogorov). We 
show how to make a completely consistent probabilistic model
by taking into account the probabilities of selecting the
settings of the PBSs. Our model matches both the experimental data and
is consistent with classical probability theory.
\end{abstract}


\section{Introduction}
We remind that Bell's type inequalities \cite{B}--\cite{Wig} are
purely probabilistic statements which a priori have no direct
relation with QM. They could be easily derived if one starts with
a Kolmogorov probability space
$$
{\cal P} =(\Omega, {\cal F}, \bf P)
$$
and a few random variables, say $A^{(i)}(\omega),
B^{(j)}(\omega).$ Here $\Omega$ is a set of chance parameters
$\omega, {\cal F}$ is a collection of subsets of $\Omega$ (so
called $\sigma$-algebra), and $\bf P$ is a probability measure.

The problem arises when one puts (by Bell's recommendation)
statistical data collected in a few experiments into those of
these inequalities which could be experimentally verified. They
are violated! (see e.g. \cite{AS}--\cite{Weihs}) Physicists
typically point out to such mystical things as non-locality or
(and) "death of realism" to explain why experimental data does not
match Bell's type inequalities,  \cite{B}--\cite{Wig}.

However, we could not ignore a possible purely probabilistic
source of violation of Bell's type inequalities. We recall that
the use of a single probability space for statistical data
collected with respect to a few different experimental contexts is
{\it not a custom of probability theory.} It is clear that if one
goes against the rules of the well established mathematical
formalism (as well as experimental statistical methodology), then
paradoxical conclusions might be expected.

We recall that Andrei Nikolaevich Kolmogorov (the founder of
modern probability theory) emphasized that each experiment is
described by its own probability space, see \cite{Kol}\footnote{
There is the evident matching between views of Kolmogorov and
Bohr. The latter pointed out that experimental arrangement should
be taken into account.}, see also Gnedenko \cite{G1}.  Thus the
use of a single probability space in the derivation of Bell's type
inequalities, see \cite{B}--\cite{Wig} for details, was not
justified. As was pointed out in \cite{FINE}--\cite{KH3} operating
with a few probability measures (corresponding to different
experiments) blocks the derivation. As was recently pointed out in
\cite{HPL2}, in probability theory such a problem -- a possibility
to realize a system of random variables on a single probability
space-  were studied and finally solved by Soviet probabilist
Vorobjev in the 1960s \cite{VR}.

If one wants to use classical probability theory to describe
the EPR-Bohm-Bell experiment, it should be done properly. To be
sure that paradoxical conclusions of violation of Bell's
inequality are not artifacts of the misuse of the mathematical
formalism, we should demand {\it Weirstrassian rigorousness of any
probabilistic description.} The aim of this paper is to provide an
alternative probabilistic model for the EPR-Bohm-Bell
experiment\footnote{ The EPR-Bohm experiment was about precise
correlations. Bell completed it by combining statistical data
collected for different experimental settings. Our point is that
this combining is responsible for paradoxical results. Therefore
we speak about "EPR-Bohm-Bell experiment".}. If one wants to apply
the classical probabilistic model, {\it a single Kolmogorov
probability space,} then random experiments for different settings
of PBS should be unified in a single random experiment in an
intelligent way. We shall describe such an experiment and we shall
show that quantum (experimental) statistical data is harmonically
combined with the classical  probabilistic description. In this
paper we shall consider the {\it Clauser-Horne-Shimony-Holt (CHSH)}
inequality, although results obtained here apply to any Bell-type
inequality.

\section{CHSH inequality}
We recall the rigorous mathematical formulation of the CHSH
inequality:

\medskip

{\bf Theorem.} {\it Let $A^{(i)}(\omega)$ and $B^{(i)}(\omega), i=
1,2,$ be random variables taking values in $[-1, 1]$ and defined
on  a single probability space ${\cal P}.$ Then the following
inequality holds:}
\begin{equation}
\label{CHSH} |<A^{(1)},B^{(1)}> + <A^{(1)},B^{(2)}> + <A^{(2)},
B^{(1)}> - <A^{(2)},B^{(2)}>| \leq 2.
\end{equation}

The classical correlation is defined as it is in classical
probability theory:
$$
<A^{(i)},B^{(j)}>= \int_\Omega A^{(i)}(\omega) B^{(j)}(\omega) d
{\bf P} (\omega).
$$
J. Bell proposed the following methodology. To verify an
inequality of this type, one should put statistical data collected
for four pairs of PBSs settings:
$$
\theta_{11}=(\theta_1, \theta_1^\prime), \theta_{12}=(\theta_1,
\theta_2^\prime), \theta_{21}=(\theta_2, \theta_1^\prime),
\theta_{22}=(\theta_2, \theta_2^\prime),
$$
into it. Here $\theta= \theta_1, \theta_2$ and $\theta^\prime=
\theta_1^\prime, \theta_2^\prime$ are selections of angles  for
orientations of respective PBSs.

Following Bell, the selection of the angle $\theta_i$ determines the
random variable
$$
A^{(i)}(\omega)\equiv a_{\theta_i}(\omega).
$$
There are two detectors coupled to the PBS with the
$\theta$-orientation: "up-spin" (or "up-polarization") detector
and "down-spin" (or "down-polarization") detector.  A click of the
up-detector assigns to the random variable $a_{\theta}(\omega)$
the value +1  and a click of the down-detector assigns to it the
value -1. However, since a lot of photons disappear without  any
click, it is also permitted for random variables to take the value
zero in the case of no detection. Therefore in Bell's framework it
is sufficient to consider $a_{\theta}(\omega)$ taking values
$-1,0,+1.$

 In the same way selection of the angle $\theta^\prime$ determines
$$
B^{(i)}(\omega)\equiv b_{\theta_i^\prime}(\omega),
$$
where $b_{\theta_i^\prime}(\omega)$ takes values $-1,0,+1.$

It seems that Bell's random model is not proper for the
EPR-Bohm-Bell experiment. Bell's description does not take into
account probabilities of choosing pairs of angles (orientations of
PBSs) $\theta_{11}, \ldots, \theta_{22}.$ Thus his model provides
only incomplete probabilistic description. This allows to include
probabilities of choosing experimental settings ${\bf P}
(\theta_{ij})$ into the model; this way completing it.

In the next section we shall provide such a complete probabilistic
description of the EPR-Bohm-Bell experiment. We point out that
random variables of our model (which will be put into the CHSH
inequality) does not coincide with Bellian variables.

\section{Proper random experiment}

\begin{itemize}
\item[a).] There is a source of entangled photons. \item[b).]
There are four PBSs and corresponding pairs of detectors. PBSs are
labelled   as $i= 1,2$ and $j=1,2$ \footnote{It is just the form
of labelling which is convenient to form pairs $(i,j).$}.
\item[c).] Directly after source there is a distribution device
which opens at each instance of time, $t=0, \tau, 2\tau, \ldots$
ways to only two (of four) optical fibers going to  the
corresponding two PBSs. For simplicity, we suppose that each pair
$(i,j): (1,1), (1,2), (2,1), (2,2)$ can be opened with equal
probability:
$$
{\bf P}(i,j)=1/4.
$$
\end{itemize}
We now define "proper random variables". To simplify
considerations, we consider the {\it ideal experiment with 100$\%$
detectors efficiency.} Thus in Bell's framework random variables
$a_{\theta}(\omega)$ and $b_{\theta^\prime}(\omega)$ should take
only values $\pm 1.$ The zero-value will play a totally different
role in our model.

\medskip

1) $A^{(i)}(\omega)= \pm 1, i=1,2$ if the corresponding (up or
down)  detector is coupled to $ith$ PBS fires;

2) $A^{(i)}(\omega)= 0$ if the $i$-th channel is  blocked. In the
same way we define random variables $B^{(j)}(\omega)$
corresponding to PBSs $j=1,2$.

\medskip

Of course, the correlations of these random variables  satisfy
CHSH inequality.

Thus if  such an experiment were performed and if CHSH inequality
were violated, we should seriously think about e.g. quantum
non-locality or death of realism.

\medskip

It would be really interesting to perform such a "proper random
experiment" for photon polarizations.

\medskip

However, to see that CHSH inequality for $<A^{(i)}, B^{(j)}>$-
correlations does not contradict to experimental data, we could
use statistical data which has been collected for experiments with
fixed pairs $\theta_{ij}= (\theta_{i}, \theta_{j}^\prime)$ of
orientations of PBS. We only need to express correlations of
Bell's variables $<a_{\theta_i}, b_{\theta_j^\prime}>$ via
correlations $<A^{(i)}, B^{(j)}>$.

\section{Frequency analysis}

Suppose that our version of  EPR-Bohm-Bell experiment was repeated
$M=4 N $ times and each pair (i,j) of optical fibers was opened
only $N$ times.

The random variables took values $$A^{(i)}= A_1^{(i)}, \ldots,
A^{(i)}_M, i=1,2, B^{(j)}= B_1^{(j)}, \ldots, B_M^{(j)}, j=1,2.$$
Then by the law of large numbers \footnote{We assume that
different trials are independent. Thus the law of large numbers is
applicable}:
$$
<A^{(i)}, B^{(j)}>= \lim_{M \to \infty} \frac{1}{M} \sum_{k=1}^M
A_k^{(i)} B_k^{(j)}.
$$
We remark that, for each pair of gates $(i,j),$ only $N$ pairs
$(A_k^{(i)}, B_k^{(j)})$ have both components non zero. Thus
$$
<A^{(i)}, B^{(j)}>= \lim_{N \to \infty} \frac{1}{4N} \sum_{l=1}^N
A^{(i)}_{k_l}B^{(j)}_{k_l},
$$
where summation is with respect to only pairs of values with both
nonzero components.

Thus the quantities $<A^{(i)}, B^{(j)}>$ are not estimates for the
$<a_{\theta_i}, b_{\theta_j^\prime}>$ obtained in physical
experiments. The right estimates are given by
$$
\frac{1}{N} \sum_{l=1}^N A^{(i)}_{k_l}B^{(j)}_{k_l}.
$$
Hence the CHSH inequality for random variables $A^{(i)}, B^{(j)}$
induces the following inequality for "traditional Bellian random
variables":
\begin{equation}
\label{CHSH1} |<a_{\theta_1}, b_{\theta_1^\prime}> +
<a_{\theta_1}, b_{\theta_2^\prime}> + <a_{\theta_2},
b_{\theta_1^\prime}> - <a_{\theta_2}, b_{\theta_2^\prime}>|\leq 8.
\end{equation}

It is not violated for known experimental data for entangled
photons. Moreover, this inequality provides a trivial constraint
on correlations: each correlation of Bellian  variables is
majorated by 1, hence, their linear combination with $\pm$-signs
is always bounded above by 4.

\section{"Proper probability space"}

We now construct a proper  Kolmogorov probability space for the
EPR-Bohm-Bell experiment. This is a general construction for
combining of probabilities produced by a few incompatible
experiments. We have probabilities $p_{ij}(\epsilon,
\epsilon^\prime), \epsilon, \epsilon^\prime= \pm 1, $ to get
$a_{\theta_i}=\epsilon, b_{\theta_{j}^\prime}=\epsilon^\prime$ in
the experiment with the fixed pair of orientations $(\theta_i,
\theta_j^\prime).$ From QM we know that
\begin{equation}
\label{QM} 
p_{ij}(\epsilon, \epsilon)=\frac{1}{2} \cos^2
\frac{\theta_i-\theta_j^\prime}{2}, p_{ij} (\epsilon,
-\epsilon)=\frac{1}{2} \sin^2 \frac{\theta_i-\theta_j^\prime}{2}.
\end{equation}

However, this special form of probabilities is not important for
us. Our construction of unifying Kolmogorov probability space
works well for any collection of probabilities $p_{ij}:
\sum_{\epsilon, \epsilon^\prime} p_{ij} (\epsilon,
\epsilon^\prime)=1.$ We remark that $p_{ij} (\epsilon,
\epsilon^\prime)$ determine automatically marginal probabilities:
$$
p_{i}(\epsilon)= \sum_{\epsilon^\prime} p_{ij} (\epsilon,
\epsilon^\prime),
$$
$$
p_{j}(\epsilon^\prime)= \sum_\epsilon p_{ij} (\epsilon,
\epsilon^\prime).
$$
In the EPR-Bohm-Bell experiment they are equal to $1/2$. Let us now
consider the  set $\{-1, 0, + 1\}^4:$ the set of all vectors
$$
\omega=(\omega_1, \omega_2, \omega_3, \omega_4), \omega_l= \pm 1,
0.
$$
It contains $3^4$ points. Now we consider the following subset
$\Omega$ of this set:
$$
\omega= (\epsilon_1, 0, \epsilon_1^\prime, 0), (\epsilon_1, 0, 0,
\epsilon_2^\prime), (0, \epsilon_2, \epsilon_1^\prime, 0), (0,
\epsilon_2, 0, \epsilon_2^\prime).
$$
It contains 16 points. We define the following probability measure
on~$\Omega:$
$${\bf P}(\epsilon_1, 0, \epsilon_1^\prime, 0) = \frac{1}{4} p_{11}(\epsilon_1, \epsilon_1^\prime),
{\bf P}(\epsilon_1, 0, 0, \epsilon_2^\prime) = \frac{1}{4}
p_{12}(\epsilon_1, \epsilon_2^\prime)
$$
$$
{\bf P}(0, \epsilon_2, \epsilon_1^\prime, 0) = \frac{1}{4}
p_{21}(\epsilon_2, \epsilon_1^\prime), {\bf P}(0, \epsilon_2, 0,
\epsilon_2^\prime) = \frac{1}{4} p_{22}(\epsilon_2,
\epsilon_2^\prime).
$$
We remark that we really have
$$
\sum_{\epsilon, \epsilon_1^\prime} {\bf P}(\epsilon_1, 0,
\epsilon_1^\prime, 0) + \sum_{\epsilon_1, \epsilon_2^\prime} {\bf
P}(\epsilon_1, 0, 0, \epsilon_2^\prime) + \sum_{\epsilon_2,
\epsilon_1^\prime} {\bf P}(0, \epsilon_2, \epsilon_1^\prime, 0) +
\sum_{\epsilon_2, \epsilon_2^\prime} {\bf P}(0, \epsilon_2, 0,
\epsilon_2^\prime) =
$$
$$
\frac{1}{4}\left[\sum_{\epsilon, \epsilon_1^\prime} p_{11}
(\epsilon_1, \epsilon_1^\prime) + \sum_{\epsilon_1,
\epsilon_2^\prime} p_{12} (\epsilon_1, \epsilon_2^\prime) +
\sum_{\epsilon_2, \epsilon_1^\prime} p_{21} (\epsilon_2,
\epsilon_2^\prime) + \sum_{\epsilon_2, \epsilon_2^\prime} p_{22}
(\epsilon_2, \epsilon_2^\prime)\right]=1.
$$
We now define random variables $A^{(i)}(\omega), B^{(j)}
(\omega):$
$$
A^{(1)}(\epsilon_1, 0, \epsilon_1^\prime, 0)= A^{(1)}(\epsilon_1,
0, 0, \epsilon_2^\prime)= \epsilon_1, A^{(2)}(0, \epsilon_2,
\epsilon_1^\prime, 0)= A^{(2)}(0, \epsilon_2, 0,
\epsilon_2^\prime)= \epsilon_2;
$$
$$
B^{(1)}(\epsilon_1, 0, \epsilon_1^\prime, 0)= B^{(1)}(0,
\epsilon_2, \epsilon_1^\prime, 0)= \epsilon_1^\prime,
B^{(2)}(\epsilon_1, 0, 0, \epsilon_2^\prime)= B^{(2)}(0,
\epsilon_2, 0, \epsilon_2^\prime)=\epsilon_2^\prime.
$$
We find two dimensional probabilities $${\bf P}(\omega \in \Omega:
A^{(1)} (\omega)= \epsilon_1, B^{(1)}(\omega)= \epsilon^\prime_1)=
{\bf P}(\epsilon_1, 0, \epsilon_1^\prime, 0)= \frac{1}{4} p_{11}
(\epsilon_1, \epsilon_1^\prime), \ldots,
$$
$$
{\bf P} (\omega \in \Omega: A^{(2)} (\omega)= \epsilon_2,
B^{(2)}(\omega)= \epsilon_2^\prime) = \frac{1}{4} p_{22}
(\epsilon_2, \epsilon_2^\prime).
$$
We also consider the random variable which is responsible for
selection of pairs of gates:
$$
\eta(0, \epsilon_2, 0, \epsilon_2^\prime)= 22, \eta(0, \epsilon_2,
\epsilon_1^\prime, 0)= 21, \eta(\epsilon_1, 0, 0,
\epsilon_2^\prime)= 12, \eta(\epsilon_1, 0, \epsilon_1^\prime, 0)=
11.
$$
It is uniformly distributed (by our assumption on equal frequency
to open each of pair of channels):

In probability theory we have the notion as {\it conditional
expectation} of a random variable (under the condition that some
event occurred).

Let $(\Omega, {\cal F}, {\bf P})$ be an arbitrary probability
space and
 let $\Omega_0 \subset \Omega, \Omega_0 \in {\cal F,} {\bf P}(\Omega_0) \ne 0.$ We also consider an arbitrary random
variable $\xi:\Omega \to {\bf R}.$ Then
$$
E(\xi|\Omega_0)= \int_\Omega \xi(\omega) d{\bf
P}_{\Omega_0}(\omega),
$$
where the conditional probability is defined by the Bayes'
formula:
$$
{\bf P}_{\Omega_0} (U) \equiv {\bf P}(U|\Omega_0)= {\bf P}(U \cap
\Omega_0)/{\bf P}(\Omega_0).
$$
Let us come back to our unifying probability space. Take
$\Omega_0\equiv \Omega_{ij}= \{ \omega \in  \Omega: \eta (\omega)=
ij\}.$ We have ${\bf P}(\Omega_{ij})= 1/4.$ Thus
$$
E(A^{(i)} B^{(j)}|\eta=ij)= \int_\Omega A^{(i)}(\omega) B^{(j)}
(\omega) d {\bf P}_{\Omega ij} (\omega) = 4\int_{\Omega_{ij}}
A^{(i)}(\omega) B^{(j)} (\omega) d {\bf P}(\omega)$$
$$
= 4\int_\Omega A^{(i)}(\omega) B^{(j)} (\omega) d {\bf P}(\omega)=
4 <A^{(i)}, B^{(j)}>= <a_{\theta_i}, b_{\theta_j^\prime}>.
$$
Thus QM-correlations for fixed choice of settings of PBSs can be
represented as conditional expectations:
\begin{equation}
\label{PI} <a_{\theta_i}, b_{\theta_j^\prime}>= E(A^{(i)}
B^{(j)}|\eta=ij).
\end{equation}

{\bf Remark.} (Jaynes critique of derivation of Bell's inequality)
Jaynes \cite{J} criticized derivation of Bell's inequality which
was based on Bell-Clauser-Horne-Shimony (CHSH) locality condition
(factorization condition). Jaynes emphasized that Bell did a
mistake in operation with conditional probabilities, because he
used the objective interpretation of probability, instead of the
subjective one. Opposite to Jaynes, we do not appeal to subjective
probability. Moreover, our aim is not critique of some special
types of derivations of Bell's type inequality. We point out that
Bell's description of the random experiment for measurement of
polarization (or spin) projections for a few incompatible pairs of
setting was incomplete. By completing this description we obtain a
classical probabilistic model which matches the experimental
data.

\section{Two-valued random variables}

We showed in the last section how to give a complete probabilistic
description of an EPR-Bohm-Bell experiment with random variables
$A^{(1)}$,$A^{(2)}$, $B^{(1)}$,  $B^{(2)}$,
and $\eta$. In that description
the $A^{(i)}$, $B^{(j)}$ took three values: $\pm 1$ and 0.
In this section we show that it is
also possible to do this when the $A^{(i)}$, $B^{(j)}$ take
only the values $\pm 1$.

By way of illustration, let us take the standard idealized
EPR-Bohm-Bell experiment described in the beginning of the previous
section with fixed orientations
$\theta_1 = \pi/4$, $\theta_2 = 0$, $\theta_1^\prime = \pi/8$,
$\theta_2^\prime = 3\pi/8$. The probabilities of the experimental
outcome
$a_{\theta_i}=\epsilon, b_{\theta_{j}^\prime}=\epsilon^\prime$ 
are given by (\ref{QM}) and yield the expected values 
\begin{equation}
\label{outcome}
<a_{\theta_1}, b_{\theta_1^\prime}> =
<a_{\theta_1}, b_{\theta_2^\prime}> = <a_{\theta_2},
b_{\theta_1^\prime}> = \frac{1}{\sqrt{2}},
 <a_{\theta_2}, b_{\theta_2^\prime}> = -\frac{1}{\sqrt{2}}
\end{equation}
Therefore we have
\begin{equation}
<a_{\theta_1}, b_{\theta_1^\prime}> +
<a_{\theta_1}, b_{\theta_2^\prime}> + <a_{\theta_2},
b_{\theta_1^\prime}> - <a_{\theta_2}, b_{\theta_2^\prime}>~=~ 2 \sqrt{2},
\end{equation}
obtaining the Tsirelson bound \cite{Ts} on the maximum quantum
"violation" of the CHSH inequality.

We construct a  Kolmogorov probability space
${\cal P} =(\Omega, {\cal F}, \bf P)$
with sixteen outcomes and five random variables:
$A^{(1)}$, $A^{(2)}$, $B^{(1)}$,  $B^{(2)}$,  $\eta$. 
The first four random variables take values $\pm 1$ and $\eta$
takes values from 11, 12, 21, 22. 
\newpage
The first eight outcomes each occur with equal
probability
$x$:
\begin{center}
\begin{tabular}{rrrrc}
$A^{(1)}(\omega)$ & $A^{(2)}(\omega)$ & $B^{(1)}(\omega)$ & $B^{(2)}(\omega)$ & $\eta(\omega)$ \\
   1 & 1 & 1 & 1 & 11 \\
   -1 & -1 & -1 & -1 & 11 \\
   1 & 1 & 1 & 1 & 12 \\
   -1 & -1 & -1 & -1 & 12 \\
   1 & 1 & 1 & 1 & 21 \\
   -1 & -1 & -1 & -1 & 21 \\
   1 & 1 & 1 & -1 & 22 \\
   -1 & -1 & -1 & 1 & 22 
\end{tabular}
\end{center}
The remaining eight outcomes each occur with equal
probability
$y$:
\begin{center}
\begin{tabular}{rrrrc}
$A^{(1)}(\omega)$ & $A^{(2)}(\omega)$ & $B^{(1)}(\omega)$ & $B^{(2)}(\omega)$ & $\eta(\omega)$ \\
   -1 & -1 & 1 & 1 & 11 \\
   1 & 1 & -1 & -1 & 11 \\
   -1 & -1 & 1 & 1 & 12 \\
   1 & 1 & -1 & -1 & 12 \\
   -1 & -1 & 1 & 1 & 21 \\
   1 & 1 & -1 & -1 & 21 \\
   -1 & -1 & 1 & -1 & 22 \\
   1 & 1 & -1 & 1 & 22
\end{tabular}
\end{center}
The probabilities $x$ and $y$ must be non-negative and $8x + 8y = 1$.
One may verify that for $i=1,2$ and $\epsilon = \pm 1$: 
$$
{\bf P} (\omega \in \Omega: A^{(i)} (\omega)= \epsilon )~=~  \frac{1}{2}.
$$
Furthermore we can check that for $i,j=1,2$ and $\epsilon = \pm 1$:
$$
{\bf P} (\omega \in \Omega: A^{(i)} (\omega)= \epsilon |
  \eta  (\omega) = i1~or~i2 ) 
= {\bf P} (\omega \in \Omega: A^{(i)} (\omega)= \epsilon 
|  \eta  (\omega) = ij)
 = \frac{1}{2}
$$
and so the {\em non-signalling} condition holds.
A similar set of equations hold for the random variables $B^{(j)}$.
We see that
$$
<A^{(i)}> = <B^{(j)}> = 0, 
$$
$$
<A^{(1)},B^{(1)}>= <A^{(2)},B^{(1)}>=8x-8y,
$$
and
$$
<A^{(1)},B^{(2)}>= <A^{(2)},B^{(2)}>=4x-4y.  
$$
The left hand side of inequality
(\ref {CHSH}) becomes $|16x-16y|$, and so (unsurprisingly)
(\ref {CHSH}) holds since $0 \leq x,y \leq 1/8$.

A further calculation shows that
\begin{equation}
\label{E1}
E(A^{(i)}B^{(j)}|\eta=ij) = 8x - 8y,~~~ij \neq 22
\end{equation}
and
\begin{equation}
\label{E2}
E(A^{(2)}B^{(2)}|\eta=22) = 8y - 8x.
\end{equation}
It suffices to set 
\begin{equation}
\nonumber
x= \frac{\sqrt{2}+1}{16\sqrt{2}},~~~y= \frac{\sqrt{2}-1}{16\sqrt{2}}
\end{equation}
in (\ref{E1}) and (\ref{E2}) to see that 
equation (\ref{PI}) is indeed satsified for the
the expected values given in
(\ref{outcome}).
Again we conclude that there is a probabilistic model consistent
with the experimental outcomes given by (\ref{outcome}).

Even more striking, perhaps, is the case when $x=1, y=0$.
From (\ref{E1}) and (\ref{E2}) we have that
\begin{eqnarray*}
\nonumber 
E(A^{(1)}B^{(1)}|\eta=11)+
E(A^{(1)}B^{(2)}|\eta=12)+~~~~~~~~~~ \\
~~~~~~~~~~E(A^{(2)}B^{(1)}|\eta=21)-
E(A^{(2)}B^{(2)}|\eta=22) = 4
\end{eqnarray*}
and so the left hand side obtains its maximum mathematical value
for any distribution of $\pm 1$ valued random variables.
Since this is larger than Tsireleson's bound of $2 \sqrt{2}$
these outcomes are not obtainable in QM. 
The above construction
gives a perfectly satisfactory probability space consistent with
these conditional expectations that
satisfies the non-signalling condition. 

{\bf Remark.} The probability space
constructed in this section 
gives values to random variables corresponding to values that are not 
measured in the EPR-Bohm-Bell experiment. 
For example, in the probablility
space 
$\omega = (1,1,1,-1,22)$ asserts that 
$A^{(1)}(\omega) = B^{(1)}(\omega)=1 $ and $\eta (\omega) = 22$.
In an EPR-Bohm-Bell experiment when the PBS's are in their second position
there are no readings for $a_{\theta_1}(\omega)$ and 
$b_{\theta_1^\prime}(\omega)$,
and QM gives no predictions about their value.
We do not assert that "in reality" for this outcome 
$a_{\theta_1}(\omega) = b_{\theta_1^\prime}(\omega) = 1$. 
After all, as pointed out,
there may be many consistent ways to assign values to 
$A^{(1)}(\omega)$ and $B^{(1)}(\omega)$. 
One interpretation of Bell's theorem is that there does not, however,
exist any such probability space consistent
with  (\ref{outcome}) for which for all i=1,2 and j=1,2:
\begin{equation}
E(A^{(i)}B^{(j)}|\eta=ij) = E(A^{(i)}B^{(j)}).
\end{equation}

We merely assert that probability spaces
exist that are consistent with all the
{\em available}
experimental data. Calculations made within the probability space
yielding formulae for which all the parameters can be measured
may be tested experimentally.

\section{Macroscopic realization of the experiment}

The experimental setting which we described in this paper, "proper
EPR-Bohm-Bell experiment", can be realized in various situations
outside quantum physics, e.g. in "classical engineering."

An example for an experiment with the outcomes described above is
the following. A device is equipped with four sensors, $A_1$,
$A_2$, $B_1$, $B_2$. Both $A$-sensors operate on a common power
supply, as do the $B$-sensors. A measurement of any of the sensors
needs the full capacity of its power supply. Thus only one of the
$A$-sensors can be active at any time and the same for the
$B$-sensors. Inactive sensors return a default value $0$. The
device randomly switches between $A_1$ and $A_2$ respectively
$B_1$ and $B_2$. When the device is polled exactly one $A$ and one
$B$ sensor return a non-default readings.

One might examine similar experimental settings outside quantum
physics. However, it seems that it would not surprise anybody from
engineering. It is not a custom to combine the data from sensors
which could not operate simultaneously.

\medskip

This paper was written during the visit of A. Khrennikov to Danish
Technical University in May 2008 and finally completed during the
visit of P. Fischer to V\"axj\"o University.

\end{document}